\documentclass[a4paper]{jpconf}
\usepackage{graphicx}
\DeclareGraphicsExtensions{.pdf,.png,.jpg,.mps,.gif}
\usepackage{indentfirst}
\usepackage{amssymb}
\usepackage{subfigure}
\usepackage[intlimits]{amsmath}
\usepackage{amsfonts}
\usepackage{textcomp}
\usepackage{rotating}
\usepackage{setspace}
\usepackage{latexsym} 
\usepackage{wasysym}
\usepackage{mathrsfs}
\usepackage{epstopdf}
\usepackage{frontespizio}
\usepackage{natbib}
\usepackage{ textcomp }
\usepackage{slashed}
\usepackage[a4paper,inner=3.cm,outer=3.9cm,top=5cm, bottom=4.1cm,pdftex]{geometry}
\usepackage{url}
 \def\mean#1{\left< #1 \right>}

\begin{document}
\title{A new approach to the variability characterization of active galactic nuclei}

\author{R Middei, F Vagnetti, M Antonucci, R Serafinelli}

\address{ Dipartimento di Fisica, Universit\`a di Roma ``Tor Vergata'', via della Ricerca Scientifica 1, 00133 Roma, Italy}

\ead{riccardo.middei@gmail.com}

\begin{abstract}The normalized excess variance is a popular method used by many authors to estimate the variability of active galactic nuclei (AGNs), especially in the X-ray band.
We show that this estimator is affected by the cosmological time dilation, so that it should be appropriately corrected when applied to AGN samples distributed in wide redshift intervals. We propose a formula to modify this estimator, based on the use of the structure function. To verify the presence of the cosmological effect and the reliability of the proposed correction, we use data extracted from the XMM-Newton Serendipitous Source Catalogue, data release 5 (XMMSSC-DR5), and cross-matched with the Sloan Digital Sky Survey quasar catalogue, of data release 7 and 12.
\end{abstract}

\section{Introduction}

Active galactic nuclei are bright sources shining both in the local and the high redshift Universe. These objects share a number of characteristic properties among which variability.
Flux variations occur in the whole electromagnetic spectrum and act on many timescales, in the case of the X-ray domain, from fraction of a day up to several years.
Studies concerning the AGN variability in the X-ray domain are of great importance in order to constrain the physics of the inner part of these sources. 
Often variability studies have been produced using the so-called normalised excess variance (NXS) estimator (e.g. [1, 2]), while only few ensemble studies based on the structure function (SF) are available, e.g. [3].
The NXS is a very popular method but it has been pointed out that it depends on the duration of the light curves [4].
We demonstrate that this method can not be used in a naive way. It is affected by the cosmological time dilation, so that, when used for a wide range of redshifts, it can lead to erroneous variability estimations.

\section{Variability estimators}

The NXS provides a straightforward way to characterize the mean variability of a single source. It is defined by the equation:
\begin{equation}
\sigma^2_{NXS}=\frac{S^2-\sigma_{n}^2}{\mean{f}^2}~,
\end{equation}
where $\mean{f}=\sum_{i=1}^Nf_i/N$ is the mean flux computed over the available flux measures $f_i$ of the same source, $S^2=\sum_{i=1}^{N}(f_i^2-\mean{f}^2)/N$ is the total variance of the light curve, while $\sigma_{n}^2$ is the mean square photometric error associated to the measured fluxes $f_i$. The $\sigma^2_{NXS}$ is the mean quadratic variation corrected for the error, so that it estimates the intrinsic variability. It can be used for approximate variability characterizations even for sources with only two measurements of the flux at different epochs. When more data are available its estimate becomes more precise (see Fig.~2).

On the other hand, the structure function analysis can be used to describe the variability of a single source as a function of the time lag between observations, but it requires a large number of data, of the order of some hundreds, to properly address the variability at the different time lags. The power of this tool is instead unveiled when the SF is used to provide ensemble variability characterizations. In both cases, the structure function describes variability as a function of the time lag $\tau$ between two observations as expressed in the formula:
\begin{equation}
SF(\tau)=\sqrt{\mean{[\log f_X (t+\tau)-\log f_X(t)]^2}-\sigma^2_{n}}~,
\end{equation}
\noindent where $\sigma_n^2$ is the mean quadratic error as in Eq.~1, and the mean can be computed on a sample of sources for ensemble analyses, or on the different flux measurements of the same source for individual studies.
It is found that the SF increases with the time lag without evidence of any flattening up to few years, and the increase can be approximated by a power law function  [3]: 
\begin{equation}
SF\propto \tau^\beta~,
\end{equation}
where $\beta=0.10\pm0.01$ has been estimated. 

As introduced in the previous section, the NXS estimator depends on the duration of the light curve. This implies that, if we compare variability of several sources using NXS we have to choose a fixed time span for all the light curves. 
When variability of sources in a wide redshift interval is investigated, light curves can still have different durations in their rest-frames. This is due to the well known effect of the cosmology, already pointed out in [5]. It leads to the relation $\Delta t_{rest}=\Delta t_{obs}/(1+z)$. This formula shows that even if in our reference frame we select only equally long light curves, they will not have the same lengths in the rest frame of the sources. 
Fig.~1 explains how the cosmological time dilation acts on the light curves, indeed, it shows how their length changes as a function of the redshift. Considering that sources have different redshift values, they all present light curves with different length in their own rest-frame.

Considering the dependence on the lag between the observations described by the SF, we expect that the NXS itself will depend on the duration of the monitoring time in the rest-frame (which for our data spans between days and years, see Sect.~3) through the integrated effect of the SF within such interval. In fact, neglecting the photometric noise both in Eq.~1 and 2, we find 
\begin{equation}
\sigma^2_{NXS}=S^2/\mean{f}^2\approx \mean{(\delta \log f_X)^2}/(\log e)^2=\mean{SF^2}/[2(\log e)^2]~,
\end{equation}
\noindent where the factor 1/2 accounts for two independent measurements contributing to each flux difference, and the average is computed over the rest-frame interval corresponding to the observed monitoring time.
Using Eq.~3, finally we find:
\begin{equation}
\sigma^2_{NXS}\propto\Delta t_{rest}^{2\beta}~.
\end{equation}

\begin{figure}
\centering
\includegraphics [width=4.4in]{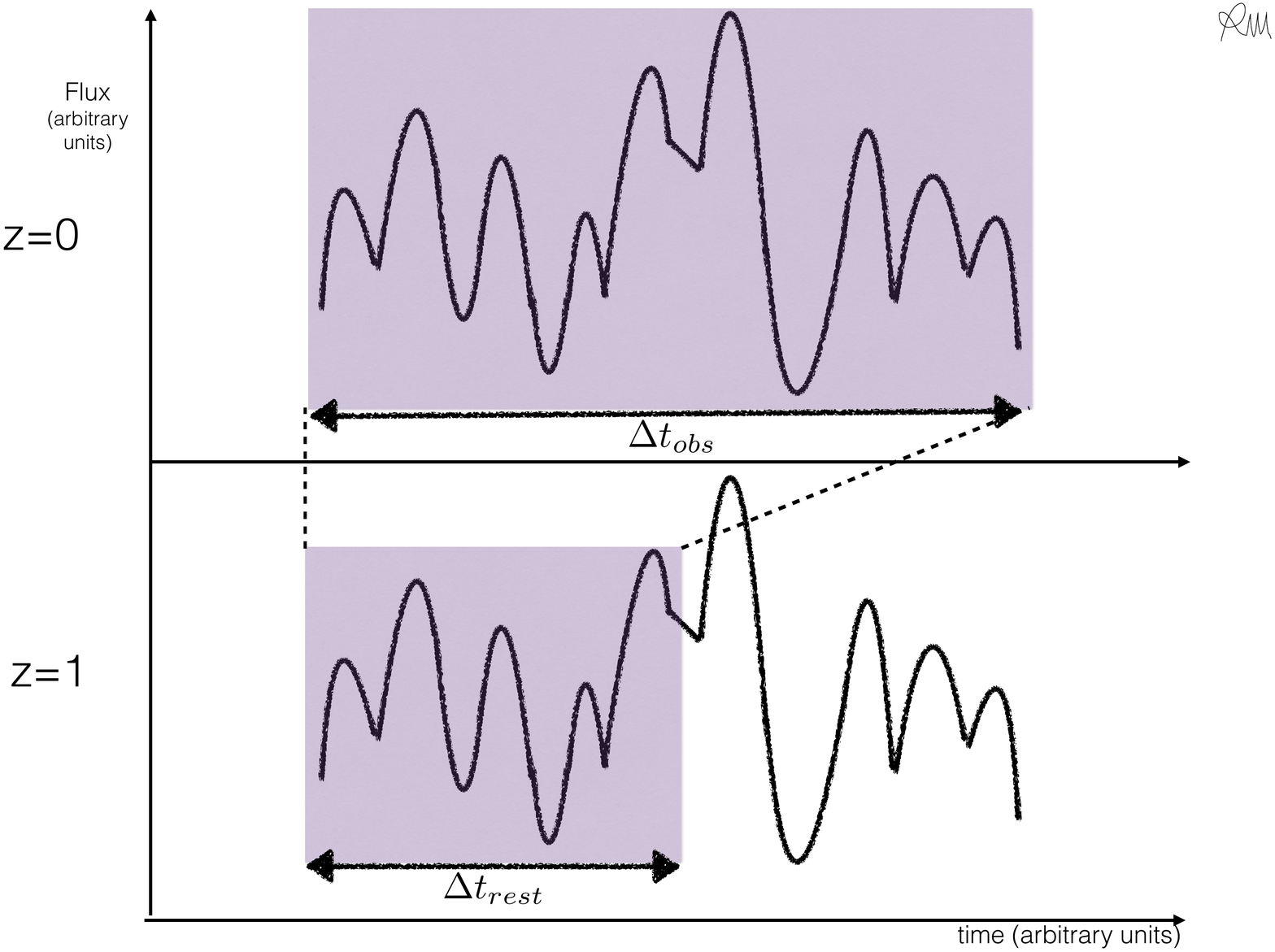}
			\caption{The effect of redshift on NXS variability estimates is shown. In the top panel it is drawn a light curve that belongs to a quasar located at $z=0$ and, for comparison in the bottom panel, that of a quasar in a farther region of the Universe ($z=1$). The intrinsic light curve is the same in both cases but, as emphasized by the shadowed area, its observed part changes. In the case of the local source it is detected a larger amplitude in flux variations that provides a greater $\sigma_{NXS}$ value. Conversely, when the object is located at high redshift, it is observed a smaller amplitude in flux variations, with a smaller value of NXS.}
	\label{fig:1}	
\end{figure}

\section{Data Sample}

Quasar variability studies are based on the comparison between at least two flux measurements of the same source, thus a suitable data set has to contain multiple observations for several sources. The XMM-Newton Serendipitous Source Catalogue (XMMSSC DR5) [6] contains a large number of X-ray detections (565,962) for 396,910 single sources achieved between 2000 and 2013. In this catalogue to each observation is associated a warning flag (SUM\_FLAG), summarizing its photometric quality. We discard all detections labelled with SUM\_FLAG$>$2 i.e. suggesting that the observation has a high probability to be artifact and unreliable.
To extract only quasar observations we take advantage of the 7th [7] and the 12th [8] data releases of the Sloan Digital Sky Survey (SDSS) quasar catalogues. These catalogues as well as the XMMSSC DR5 contain  for each observation the equatorial coordinates, $\alpha$ and $\delta$. This allows us to cross-match the catalogues by coordinates, thus to extract only quasar data from the serendipitous X-ray detections.
We consider a match when an XMMSSC observation lies within 5 arcsec in both $\alpha$ and $\delta$ from an observation listed in any of the two SDSS quasar catalogues.
\begin{figure}
\centering
		\includegraphics [width=4.5in]{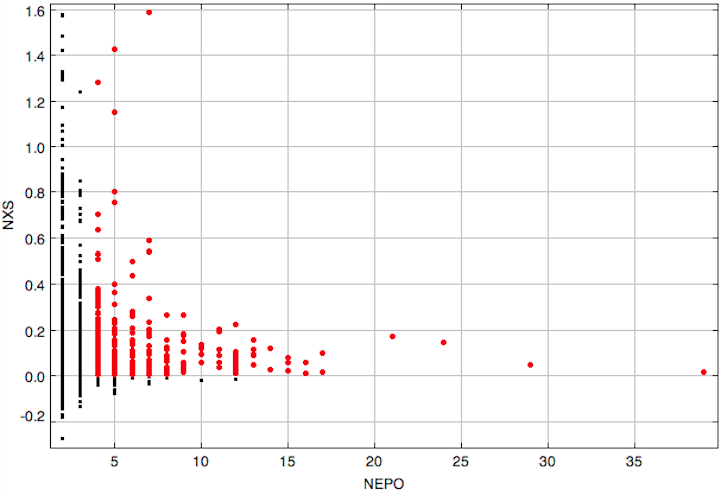}
			\caption{NXS estimates of our sources as a function of the number of epochs (NEPO). Points represent the NXS values computed. Notice that when the NXS parameter is computed using a larger number of measurements its values are distributed in a narrower region. Black points refer to NXS estimates computed with less than four observations and/or to negative values for NXS. Red points constitute the subsample we use in the analysis and are characterized by positive NXS values obtained with at least four observations.}
	\label{fig:1}	
\end{figure} 

The joint data set contains only spectroscopically confirmed quasars, and as a further step we require that sources have at least two observations. 
This leads us to obtain a preliminary sample of 2,700 sources (all the points in Fig.~2). Referring to Eq.~1, it might be possible to obtain $\sigma_{NXS}^2<0$ when the mean squared error is larger than the variance. In such cases we discard the NXS estimates, because the variability information provided is embedded into the noise, so reducing the sample to 1,775 sources. Furthermore, since the NXS parameter is an averaged estimate of the source variability, it is desirable to increase the number of measurements used to compute NXS. The computed $\sigma_{NXS}^2$ will better approximate the intrinsic NXS of the source when a large number of data is used. However, only few sources are observed several times so that, in this work we used a subsample containing 416 sources observed at least four times, the red points in Fig.~2.

\section{Trend and correction}

We report in Fig.~3 (blue points) the NXS estimates computed according to Eq.~1.
The linear least squares fit to these estimates shows how the cosmic time dilation acts on the NXS, yielding a $\sigma_{NXS}^2\propto \Delta t_{rest}^{0.21\pm0.04}$. The Pearson correlation coefficient has a value $r=0.24$, while the probability to find by chance the same correlation is $P(>r)=6\times10^{-7}$.
\begin{figure}[h]
	\centering
		\includegraphics [width=4.4in]{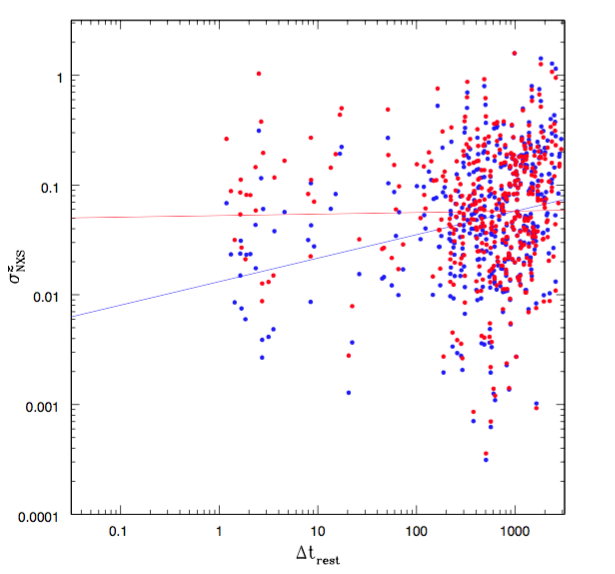}
			\caption{Comparison of the corrected (red) and uncorrected (blue) NXS estimates and their least squares fits as functions of the rest-frame light curve duration $\Delta t_{rest}$.}
	\label{fig:1}	
\end{figure} 
It is possible to correct the NXS estimates using the ensemble description provided by the structure function. The SF ensemble analysis, indeed, characterizes the variability of a large number of sources as a function of the lag $\tau$ between two observations. Eq.~5 describes how $\sigma_{NXS}^2$ depends on the duration of the light curve in the rest-frame, $\Delta t_{rest}$, and it allows us to extrapolate the expected variability for a certain source at a certain time lag. In order to compare the variability of sources characterized by shorter light curves with those objects having longer light curves, we normalize Eq.~5 to a given time interval $\Delta t^*$ that can be freely chosen:
\begin{equation}
\sigma_{NXS}^{2~*}=\sigma^2_{NXS}\left(\frac{\Delta t^*}{\Delta t_{rest}}\right)^{2\beta}=\sigma^2_{NXS}\left(\frac{\Delta t^*}{\Delta t_{obs}}\right)^{2\beta}(1+z)^{2\beta}~.
\end{equation}
We adopt $\Delta t^*=1000$ days, and the results are shown in Fig.~3 (red points).

 Applying the correction described in Eq.~6, the formal fit to the $\sigma_{NXS}^{2~*}$ values has slope $a\sim0.001$ and Pearson correlation coefficient $r=0.06$, with $P(>r)=0.83$, indicating that the two quantities are not anymore correlated.
 
The correction prevents the NXS wrong estimates for high redshift sources, whose shorter $\Delta t_{rest}$ can be extrapolated to the same $\Delta t^*$ as for the other sources actually measured at such long times.

\section{Final Remarks}

The NXS method provides a reliable variability description only when applied to local sources, otherwise it leads to variability underestimates. This is due to the cosmological time dilation that shrinks the duration of the light curves. The correction we propose takes advantage of the statistical characterization provided by ensemble SF-based studies. With this correction it is possible to extrapolate the NXS estimates to a fixed rest-frame duration of the light curve $\Delta t^*$, so obtaining corrected values which can be compared to each other.

\section*{Acknowledgements}
We thank the organisers of the sixth Young Researchers Meeting.
We thank the Gran Sasso Science Institute for hospitality.
This research has made use of data obtained from the 3XMM XMM-Newton serendipitous source catalogue compiled by the 10 institutes of the XMM-Newton Survey Science Centre selected by ESA. 

\medskip
\medskip

\section*{References}
\medskip
\noindent $[1]$ K. Nandra, I. M. George, R. F. Mushotzky, et al. 1997, ApJ, \textbf{476}, 70\\
\noindent$[2]$ G. Ponti, I. Papadakis, S. Bianchi, et al. 2012, A\&A, \textbf{542}, A83\\
\noindent$[3]$ F. Vagnetti,  S. Turriziani, \& D. Trevese,  2011, A\&A, \textbf{536}, A84\\
\noindent$[4]$ A. Lawrence, \& I. Papadakis, 1993, ApJ, \textbf{414}, L85\\
\noindent$[5]$ C. M. Gaskell, 1981, Astrophys. Lett., \textbf{21}, 103\\
\noindent$[6]$ S. R. Rosen, N. A. Webb, M. G. Watson, et al. 2015, \verb|ArXiv: 1504.07051|\\
\noindent$[7]$ D. P. Schneider, G. T. Richards, P. B. Hall, et al. 2010, AJ, \textbf{139}, 2360\\
\noindent$[8]$ I. Paris, P. Petitjean, E. Aubourg, et al. 2014, A\&A, \textbf{563}, A54\\
\end{document}